\documentstyle[psfig]{elsart}

\begin{document}
\begin{frontmatter}

\title{Thermal excitation of heavy nuclei with 5-15 GeV/c antiproton,
  proton and pion beams}

\author[iucf]{L. Beaulieu}
\author[iucf]{K. Kwiatkowski\thanksref{LANL}}
\author[iucf]{W.-c. Hsi}
\author[iucf]{T. Lefort}
\author[war]{L. Pienkowski}
\author[sfu]{R.G. Korteling}
\author[iucf]{G. Wang\thanksref{EPS}}
\author[anl]{B. Back}
\author[iucf]{D.S. Bracken\thanksref{LANL}}
\author[mary]{H. Breuer}
\author[iucf]{E. Cornell\thanksref{LBNL}}
\author[tamu]{F. Gimeno-Nogues}
\author[iucf]{D.S. Ginger\thanksref{CAM}}
\author[bnl]{S. Gushue}
\author[msu]{M.J. Huang}
\author[tamu]{R. Laforest\thanksref{BAR}}
\author[msu]{W.G. Lynch}
\author[tamu]{E. Martin}
\author[lanl]{K.B. Morley}
\author[bnl]{L.P. Remsberg}
\author[tamu]{D. Rowland}
\author[tamu]{E. Ramakrishnan}
\author[tamu]{A. Ruangma}
\author[msu]{M.B. Tsang}
\author[iucf]{V.E. Viola}
\author[tamu]{E. Winchester}
\author[msu]{H. Xi}
\author[tamu]{S.J. Yennello}

\address[iucf]{Department of Chemistry and IUCF, Indiana University,
  Bloomington, IN 47405}
\address[war]{Heavy Ion Laboratory, Warsaw University, Warsaw Poland}
\address[sfu]{Department of Chemistry, Simon Fraser University, Burnaby,
BC, Canada V5A IS6}
\address[anl]{Physics Division, Argonne National Laboratory, Argonne IL 60439 }
\address[mary]{Department of Physics, University of Maryland, College Park, MD
20740}
\address[tamu]{Department of Chemistry \& Cyclotron Laboratory, Texas A\&M
University, College Station, TX 77843}
\address[bnl]{Chemistry Department, Brookhaven National Laboratory, Upton, NY
11973}
\address[msu]{Department of Physics and NSCL, Michigan State University, East
Lansing, MI 48824}
\address[lanl]{Los Alamos National Laboratory, Los Alamos, NM 87545}

\thanks[LANL]{Present address: Los Alamos National Laboratory, Los Alamos, NM 87545}
\thanks[EPS]{Present address: Epsilon, Inc., Dallas, TX 75240}
\thanks[LBNL]{Present address: Lawrence Berkeley Laboratory, Berkeley, CA 94720}
\thanks[CAM]{Cambridge University, Cambridge, U.K.}
\thanks[BAR]{Barnes Hospital, Washington University, St. Louis, MO  63130}

\begin{abstract}
Excitation-energy distributions have been derived from measurements of
5.0-14.6 GeV/c antiproton, proton and pion reactions with $^{197}$Au
target nuclei, using the ISiS 4$\pi$ detector array.  The maximum
probability for producing high excitation-energy events is found for
the 8 GeV/c antiproton beam relative to other hadrons, $^3$He and $\bar{p}$
beams from LEAR.  For protons and pions, the excitation-energy 
distributions are nearly independent of hadron type and beam momentum 
above about 8 GeV/c.  The excitation energy enhancement for $\bar{p}$ 
beams and the saturation effect are qualitatively consistent with 
intranuclear  cascade code predictions.  For all systems studied, 
maximum cluster sizes are observed for residues with E*/A $\sim$ 
6 MeV.\\

{\small{{\it PACS:}25.70.Pq,25.43.+t,25.80.Hp}}
\end{abstract}

\end{frontmatter}

Much effort in nuclear physics has been devoted to the study of
the formation of multiple complex fragments (3$\leq Z \leq$16), 
or multifragmentation, and its possible link to the nuclear liquid-gas 
phase transition~\cite{mor93}. In particular, the strong interest created by 
the measurement of a latent heat by Pochodzalla {\it {et al.}}~\cite{Pochodzalla95}, 
which would signal a first order phase transition. However, in many cases 
multifragmentation of heavy nuclei is not only driven by the its 
thermal and Coulomb properties, but also by collective (dynamical) 
properties of the chaotic systems formed in energetic projectile-target 
interactions. 

The thermal features of the breakup process are isolated most 
transparently in reactions induced by hadron and light-ion beams at 
energies in excess of 2 GeV\cite{Goldenbaum96,Kwiat98,Morley96}. Transport
calculations predict that such beams heat nuclei rapidly, $\tau
\leq$ 30-40 fm/c, at the same time producing little
compression, low average angular momentum and minimal shape
distortions\cite{Cu,Wang96}.  While excitation energy deposition $E^*$
in GeV hadron-induced reactions is significantly less than the total
available energy\cite{Goldenbaum96,Kwiat98}, transport calculations
indicate that residues can be created in such collisions with
$E^*$ values well in excess of the multifragmentation threshold,
$E^*/A \sim$ 5 MeV\cite{Bondorf95Botvina90,Friedman90,Gross90}.  The
objective of the present study is to investigate the relative
effectiveness of various hadron beams and momenta for producing E*
values in excess of $\sim$ 5 MeV/A, thereby identifying the optimum
system for studies of thermal multifragmentation and underlying
phenomena e.g. the liquid-gas phase transition.

In Fig.\ \ref{fig1}, excitation-energy predictions of the
Toneev\cite{Toneev90} intranuclear cascade calculation (INC) are shown.
Here the average excitation energy imparted to a $^{197}$Au nucleus by
proton, $\pi^-$ and antiproton beams is plotted as a function of beam
momentum.  For p and $\pi^-$ beams, the $\left< E^* \right>$ values
are predicted to be nearly identical; above about 8 GeV/c there is 
little dependence on beam momentum.  These features have been verified 
qualitatively in charged-particle multiplicity studies by 
Hsi {\it et al.}\cite{Hsi97} and in earlier inclusive studies by 
Porile\cite{Porile89}.  However, the antiproton
predictions exhibit a significant increase in average
excitation energy.  This enhancement derives from reabsorption of some
fraction of the annihilation pions ($< n_{\pi} > \sim$ 5), which
complements the internal heating caused by the cascade of
hadron-hadron collisions and $\Delta$ resonance
excitations\cite{Cu,Toneev90,Strottman84}. From the point of view of
multifragmentation, this greatly enhances the probability for forming
residues excited above the multifragmentation threshold, as shown in 
the inset of Fig. 1, where $E^*$ distributions for 8
GeV/c $\pi^-$ and $\bar{p}$ are compared.

In this letter we present results for excitation-energy distributions
derived from bombardments of $^{197}$Au nuclei with 5.0-14.6 GeV/c
hadron beams.  Exclusive charged-particle multiplicities and energy
spectra were measured at the Brookhaven AGS accelerator with the
Indiana Silicon Sphere, a 4$\pi$ detector array with 162
gas-ion-chamber/silicon/CsI telescopes\cite{Kwiat95b}.  Two
experiments were performed.  The first used untagged negative beams
(largely $\pi^-$) at 5.0, 8.2, and 9.2 GeV/c and positive beams
(primarily protons) at 6.2, 10.2, 12.8 and 14.6 GeV/c.  The second was
performed with a tagged 8.0 GeV/c negative beam, which provided
simultaneous measurement of the $\pi^-$ (98\%) and $\bar{p}$ (1\%) reactions.  
The results at 8 GeV/c for the two $\pi^-$ experiments are found to be
identical within error bars.  Further experimental details can be
found in\cite{Hsi97,Kwiat95b,Lefort99}.  As part of our analysis, we
have also compared with data from the 4.8 GeV $^3$He + $^{197}$Au
reaction\cite{Kwiat98} at LNS Saclay and the 1.2 GeV/c $\bar{p}$ +
$^{197}$Au reaction from LEAR\cite{Goldenbaum96}.

In the experiments reported here, calculation of the excited residue
charge and mass was made by subtracting fast cascade particles from
the target charge and mass using the same procedure as in
ref~\cite{Kwiat98}.  Excitation-energy reconstruction was performed
for each event by calorimetry according to the following
prescription:

\begin{equation}
E^* = \sum_{i=1}^{M_c} K_i + M_n <K_n> + Q + E_{\gamma}.
\end{equation}

\noindent Here, $K_i$ is the kinetic energy of each charged particle
in an event of multiplicity $M_c$, $M_n$ and $\left< K_n \right>$ are
the multiplicity and average kinetic energy of neutrons, $Q$ is the
mass difference of the reconstructed event, and $E_{\gamma}$ is a small
term to account for gamma de-excitation of the residual nucleus and
excited fragments.  This procedure is similar to those employed
in\cite{Kwiat98,Lefort99,Ste89,Pochodzalla95,Ma97,Beaulieu99,Hauger96} 
and a full paper on the dependence of $E^*$ values on the various
assumptions of the reconstruction and the effects of fluctuations is
in preparation.

For the present measurements the two most sensitive parameters of the
reconstruction procedure involve the assumption concerning $\left< K_n
\right>$ and the definition of the energy range for thermal-like
particles.  Since neutrons are not measured in ISiS, we use the
neutron-charged particle correlations reported for LEAR data
by Goldenbaum {\it et al.}\cite{Goldenbaum96}.  This correlation is in
good agreement with similar results from heavy-ion
reactions\cite{Rochester} and model
simulations\cite{Bondorf95Botvina90,Friedman90,Durand}. However,
systematic uncertainties arise when the kinetic energies are assigned
to the neutrons. The neutron average kinetic energies were taken from
the predicted correlation between $\langle K_n \rangle$ and E*/A by
SMM\cite{Bondorf95Botvina90}. Then Eq. 1 is iterated to obtain self-consistency.  
Similar values are obtained from an iterative procedure using $E^*$ = $aT^2$
and an initial value of $\langle K_n\rangle$ = $T$. Comparison between 
unfiltered and filtered simulations with SMM\cite{Bondorf95Botvina90} 
and the evaporation code SIMON~\cite{Durand} both show that the use of 
$3T/2$ (initial $\langle K_n\rangle$ = $3T/2$) overpredicts the values of 
$\langle K_n\rangle$ by as much as 30\% at high $E^*$, resulting in 
an overestimation of $E^*$ of about 10-12\% ($\sim$ 20\% for  
initial $\langle K_n\rangle$=$2T$). 

Thermal-like charged particles are defined by the spectral shapes, 
from which an upper cutoff of 30 MeV for H and $9Z+40$ MeV for heavier 
fragments was assigned~\cite{Kwiat98,Morley96}. Our definition of $E^*$ 
is a conservative one; for example, with the expanded charged particle
acceptance ($E/A <$ 30 MeV) of Hauger {\it et al.}\cite{Hauger96}, 
we obtain $E^*$ values about 25\% higher than reported here.

Finally, the stability of the $E^*$ reconstruction procedure has been 
tested using SMM and SIMON. Both models give a strong linear correlation 
between the unfiltered and filtered $E^*$. The average values are
recovered by the method with deviation not larger than $\pm$10\% over
the useful range of the data ($E^*/A$=2-9 MeV). A detailed comparison will be 
presented in a long paper. However, the approach taken above should be 
viewed as a conservative one, as should the $E^*$ distributions.

In Figure 2 we show the reconstructed probability distributions for 
excitation energy and residue mass for the range of systems studied in 
this work, where $\sum_i P(E^*_i) = 1$.  Values of $E^*$ below 250 MeV 
become highly uncertain due to the dominance of neutron emission 
(unmeasured in ISiS) at low excitation energies and the
suppression of $M_c \leq$ 2 events by the ISiS trigger.  Data for 6.2
GeV/c and 12.8 GeV/c protons (not shown) are similar to other proton 
and pion data in Figs. 2-5. Fig.\ref{fig2} demonstrates that the largest 
population of high excitation energy events is achieved with the 
8.0 GeV/c $\bar{p}$ beam and the lowest with the 5.0 GeV/c $\pi^-$ beam. 
The 12.8 GeV/c proton distribution is slightly higher than for 14.6 GeV/c p, 
while that for the 6.2 GeV/c protons is slightly lower than the 8 GeV/c $\pi^-$.  
Thus, the data and the INC predictions of Fig.\ \ref{fig1} are in qualititive 
agreement.  Quantitatively, however, the INC calculations predict $E^*$ 
distributions that extend significantly beyond the data, as discussed 
in\cite{Lefort99}.  

The residue mass
distributions show a somewhat different pattern.  In this case the
14.6 GeV/c proton beam produces the lightest residues and the 5.0 GeV/c
$\pi^-$ the heaviest.  This mass dependence can be understood as a 
consequence of the fast cascade, which produces an increasing 
number of fast knockout particles as the beam momentum increases, and 
also leads to a density-depleted residue\cite{Cu,Wang96}. This process 
produces the saturation in average excitation energy shown in the
INC calculations of Fig.\ref{fig1} and the data in Fig.\ref{fig2}. 
That is, the increase in total available beam energy for
$E^*$ deposition is counterbalanced by loss of energy due to mass loss
$\Delta A$ during the fast cascade.  This mass loss, derived
from the data, is shown in the top panel of Fig.\ref{fig3} as a 
function of deposited excitation energy.

The relative effectiveness of various beams in depositing high excitation
energies is shown in the bottom panel of Fig. \ref{fig3} and
summarized in Table 1.  Included here are comparable data from the 4.8
GeV $^3$He + $^{197}$Au reaction\cite{Kwiat98} and the 1.2 GeV
$\bar{p}$ + $^{197}$Au reaction\cite{Goldenbaum96}.   
In order to emphasize the probability for forming highly excited
systems, we examine the ratio of total events with $E^*$ greater than a
given value to that for events with $E^* \geq$ 400 MeV. At $E^*$=400 MeV,
the event reconstruction should provide the greatest self-consistency
among the data sets.

Figure \ref{fig3} and Table\ \ref{table1} confirm that the 8.0 GeV/c
antiproton beam produces a significant enhancement of high excitation
energy events, particularly in the multifragmentation regime above
800-1000 MeV.  In Table \ref{table1}, the yield of events with
excitation energy above the multifragmentation threshold for Au-like
nuclei (a range that spans 800-1000 MeV or about 5 MeV/nucleon) is
listed, compared to total events above $E^* >$400 MeV ($E^*/A >$2 MeV). The
enhancement for the 8 GeV/c $\bar{p}$ beam is approximately 25\% greater than 
the next most effective beam, 12.8 GeV/c protons. Relative to the $\bar{p}$
studies with 2.1 GeV/c $\bar{p}$ at LEAR~\cite{Goldenbaum96}, where 
negligible multifragmentation yield was observed, the probabilities for
high $E^*$ events at 8 GeV/c are over an order of magnitude greater.
In this regard, we note that for the 1.8 GeV $^3$He + $^{197}$Au 
system~\cite{Ren91}, for which the charged-particle multiplicity data are 
very similar to those with LEAR beams~\cite{Gol99}, the total cross section 
for events with four or more Z $\geq$ 3 fragments is only 3.5 mb.  At the 
higher energy of 4.8 GeV, this cross section has grown to 83 mb.  Thus, 
when account is made for the rapid growth of multifragmentation cross section 
with increasing beam momentum, the ISiS results and those of 
Ref.~\cite{Goldenbaum96} appear to be self-consistent.

In Fig. \ref{fig4} the excitation energy distributions are plotted
as a function of $E^*/A$ of the residue.  We note here, as well as in Fig~\ref{fig5},
that events with $E^*/A \geq$ 9 MeV comprise less than 1\% of the data
set. The same general features
persist as in Fig.\ \ref{fig3}, except in this case the 14.6 GeV/c
proton beam yields comparable probabilities in the region beyond about
$E^*/A >$ 9 MeV.  Two factors account for this.  First, the average residue
mass is lighter for reactions at this momentum, as shown in Fig.\ \ref{fig2}.
Second, the number of events obtained with the 8.0 GeV/c $\bar{p}$
beam ($\sim$ 25,000) were about two orders of magnitude lower than for
the other beams, creating larger statistical uncertainties at the
extremes.

Finally, in Fig.\ \ref{fig5} we examine the dependence of the fragment
size distributions on E*/A, of relevance to discussions of critical
phenomena and a nuclear liquid-gas phase
transition\cite{Porile89,Hauger96}.  The top panel shows the number of
observed IMFs per residue nucleon and the corresponding filter-corrected 
value as a function of $E^*/A$ for the various reactions.  
This ratio is nearly identical for all systems, increasing systematically 
with increasing $E^*/A$ up to 9 MeV and then becoming roughly constant
thereafter. The same uniformity is observed in
the fragment charge distributions, shown in the lower panel of Fig.\
\ref{fig5},  where the parameter $\tau$ from power-law fits to the
charge distributions are plotted as a function of $E^*/A$.  Values of
$\tau$ decrease steadily as the system is heated i.e. the
probability for forming larger fragments increases. A minimum
is reached at $\tau \sim$ 2 near $E^*/A \sim$ 6 MeV, followed by a slight 
increase (smaller fragments).  This signifies that maximum cluster sizes
are obtained very near the multifragmentation threshold.  Thereafter,
additional excitation appears to produce a hotter environment, leading
to an increased yield of lighter particles and clusters.

In summary, the heat content ($E^*$) of equilibrium-like
heavy residues formed in 5-15 GeV/c hadron-induced
reactions has been investigated.  The antiproton beam is found to be
most effective in creating highly excited residues, in qualitative 
agreement with INC predictions.  Relative to the threshold for 
multifragmentation in such systems ($E^* \sim$ 800-1000 MeV), the 
enhancement of high excitation energy events with antiprotons is
at least 25-35\% greater than other hadrons and over an order of magnitude greater
than antiprotons from LEAR.  Above momenta of about 8 GeV/c the
probability for $E^*$ deposition with hadron beams is nearly independent
of hadron type or beam momentum, again consistent with INC
calculations. The observed average number of IMF per residue 
nucleon and the power-law fits to the charge distributions show a universal 
behavior as a function of $E^*/A$. This independence of the final multifragmentation
state on collision dynamics suggests equilibrium-like behavior in the
breakup of the hot residues~\cite{Bea96,Sch96}. Therefore
hadron, especially antiprotons around 6-8 GeV/c and pions or protons above
10-12 GeV/c, are very well suited to study thermal multifragmentation. For
a given beam momentum, they provide a wide range of thermal energy ($E^*/A$), 
which is an essential quantity for investigating latent heat in
nuclear matter and related properties.

{\bf{Acknowledgements}}

The authors thank J. Vanderwerp, W. Lozowski, K. Komisarcik and
R.N. Yoder at IUCF and P. Pile, H. Brown, W. McGahern, J. Scaduto,
L. Toler, J. Bunce, J. Gould, R. Hackenburg and C. Woody at AGS for
their assistance with these experiments. This work
was supported by the U.S. Department of Energy and National Science
Foundation, the National Sciences and Engineering Research Council of
Canada, Grant No. P03B 048 15 of the Polish State Committee for
Scientific Research, Indiana University Office of Research and the
University Graduate School, Simon Fraser
University and the Robert A. Welch Foundation.


%
%

\newpage

\begin{figure}
\caption{Intranuclear cascade predictions~\protect\cite{Toneev90} of the average excitation
  energy for events with E* $>$ 50 MeV are shown as a function of momentum for p, $\pi^-$ and $\bar{p}$ beams
  incident on $^{197}$Au.  Inset compares the excitation energy
  probability distributions for 8 GeV/c $\pi^-$ and $\bar{p}$ beams}
\label{fig1}
\end{figure}

\begin{figure}
\caption{Excitation energy (left frame) and residue mass probability
  (right frame) for several of the systems studied in this work, as
  indicated on figure.}
\label{fig2}
\end{figure}

\begin{figure}
\caption{Bottom:  the probability for observing events with excitation
  energy greater than E* $\geq$ 400 MeV relative to the probability
  for events with E* = 400 MeV.  Systems are indicated on figure. Top:
  average mass loss $\Delta$A in the fast cascade as a function of
  excitation energy.  Systems are defined in bottom frame.}
\label{fig3}
\end{figure}

\begin{figure}
\caption{Probability distributions for data in Fig. 2, plotted as a
  function of E*/A of the residue.  Systems are defined on figure.}
\label{fig4}
\end{figure}

\begin{figure}
\caption{Top:  average ratio of observed and geometry corrected IMFs per residue 
  nucleon as a function of E*/A; symbols are defined in bottom frame.  Bottom:
  power-law parameters $\tau$ from fits to the charge distributions as
  a function of E*/A of the residue.  Systems are defined on figure.}
\label{fig5}
\end{figure}

%
%

\newpage

 \begin{table}
 \caption{Ratio of the integrated events beyond multifragmentation
threshold to total events with $E^* >$ 400 MeV ($E^*/A >$ 2 MeV).}
 \label{table1}
 \begin{tabular}{ccccccc}
beam & p(GeV/c) & T(GeV) & ${{P(E^* > 800 MeV)} \over {P(E^* >
400 MeV)}}$  & ${{P(E^* > 1000 MeV)} \over {P(E^* > 400 MeV)}}$  
& \multicolumn{1}{c}{ ${ {P(E^*/A > 5 MeV)} \over {P(E^*/A >
2 MeV)} }$} \\ \hline
E900a $\bar{p}$ & 8.0 & 7.2 & 0.30 & 0.097 & 0.27 \\ \hline
E900 $p$ & 14.6 & 13.7 & 0.23 & 0.067 & 0.21 \\
E900 $p$ & 12.8 & 11.9 & 0.25 & 0.076 & 0.22 \\
E900 $p$ & 10.2 & 9.3 & 0.23 & 0.066 & 0.19 \\
E900 $\pi^-$ & 9.2 & 9.1 & 0.21 & 0.058 & 0.17 \\
E900a $\pi^-$ & 8.0 & 7.9 & 0.21 & 0.056 & 0.18 \\
E900 $\pi^-$ & 8.2 & 8.1 & 0.20 & 0.054 & 0.17 \\
E900 $p$ & 6.2 & 5.3 & 0.19 & 0.045 & 0.13 \\
E900 $\pi^-$ & 5.0 & 4.9 & 0.17 & 0.036 & 0.11 \\
$^{3}$He [10]& 7.6 & 4.8 & 0.12 & 0.020 & N/A \\
$\bar{p}$ [8]& 2.1 & 1.2 & 0.042 & 0.003 & N/A 
 \end{tabular}
 \end{table}

\newpage

\begin{figure}
\centerline{\psfig{file=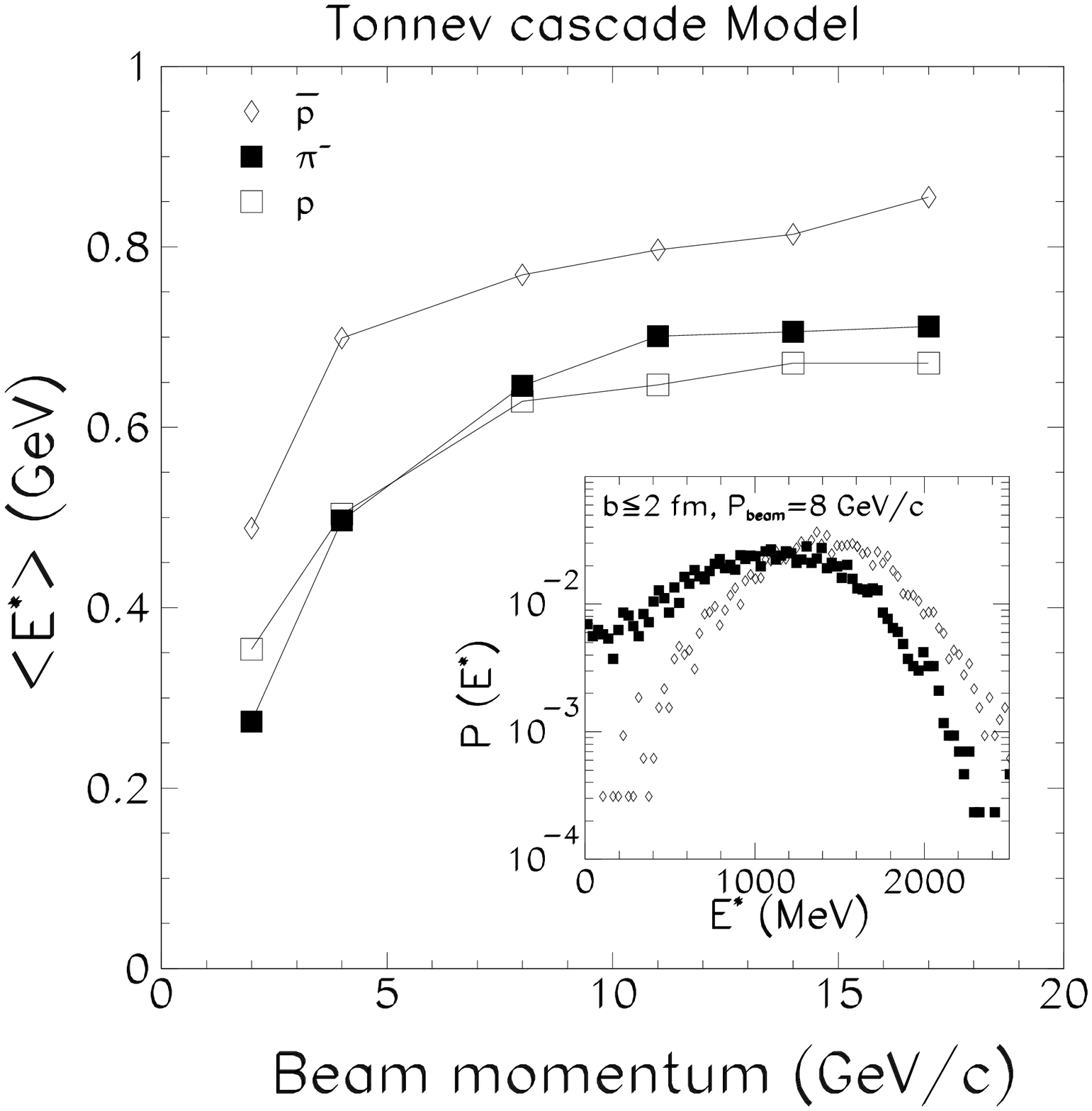,width=7.5in}}
\vspace{1in}
\center{Fig. 1: L. Beaulieu {\em et al.}}
\end{figure}

\newpage

\begin{figure}
\centerline{\psfig{file=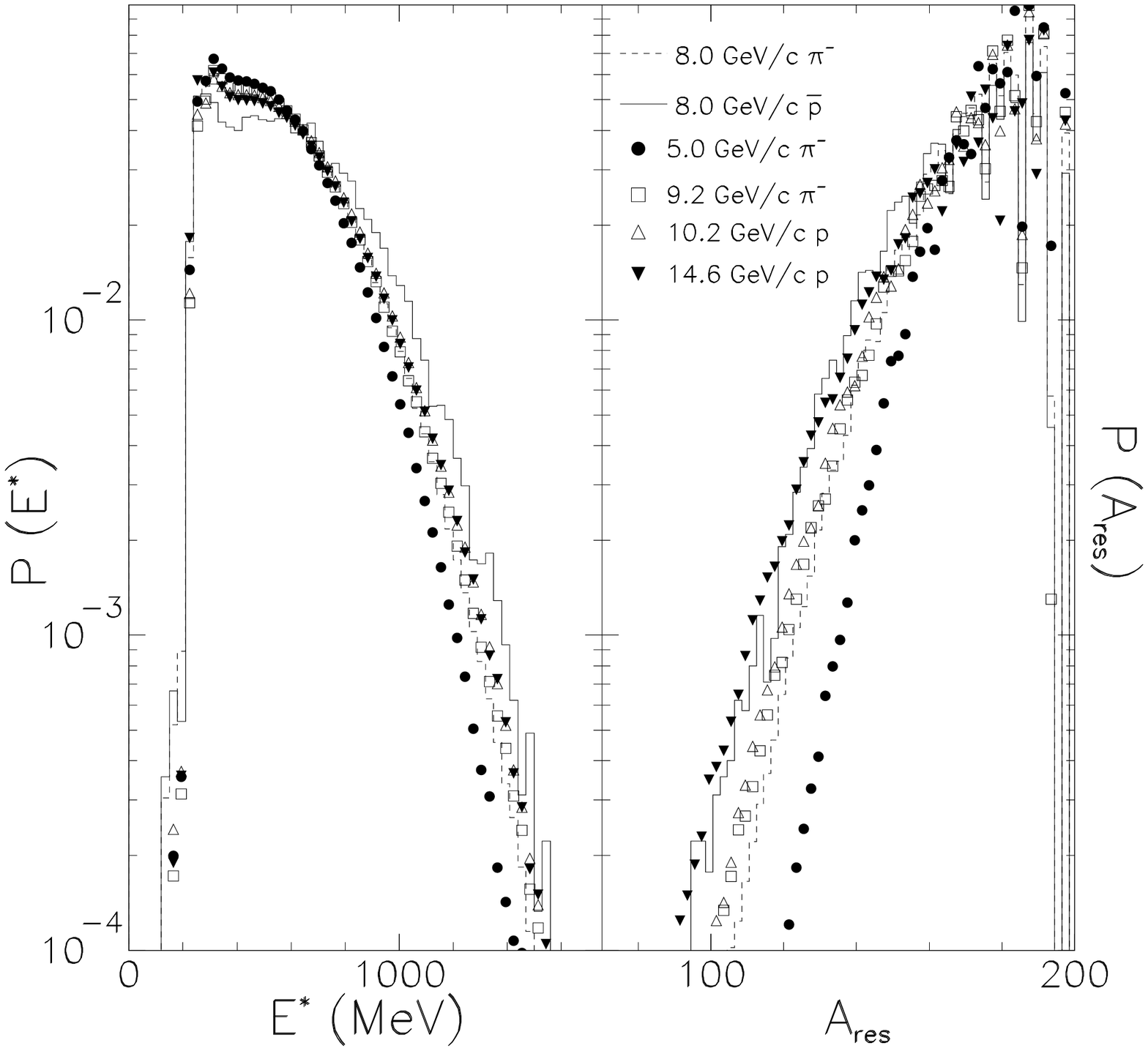,width=7.5in}}
\vspace{1in}
\center{Fig. 2.: L. Beaulieu {\em et al.}}
\end{figure}

\newpage

\begin{figure}
\centerline{\psfig{file=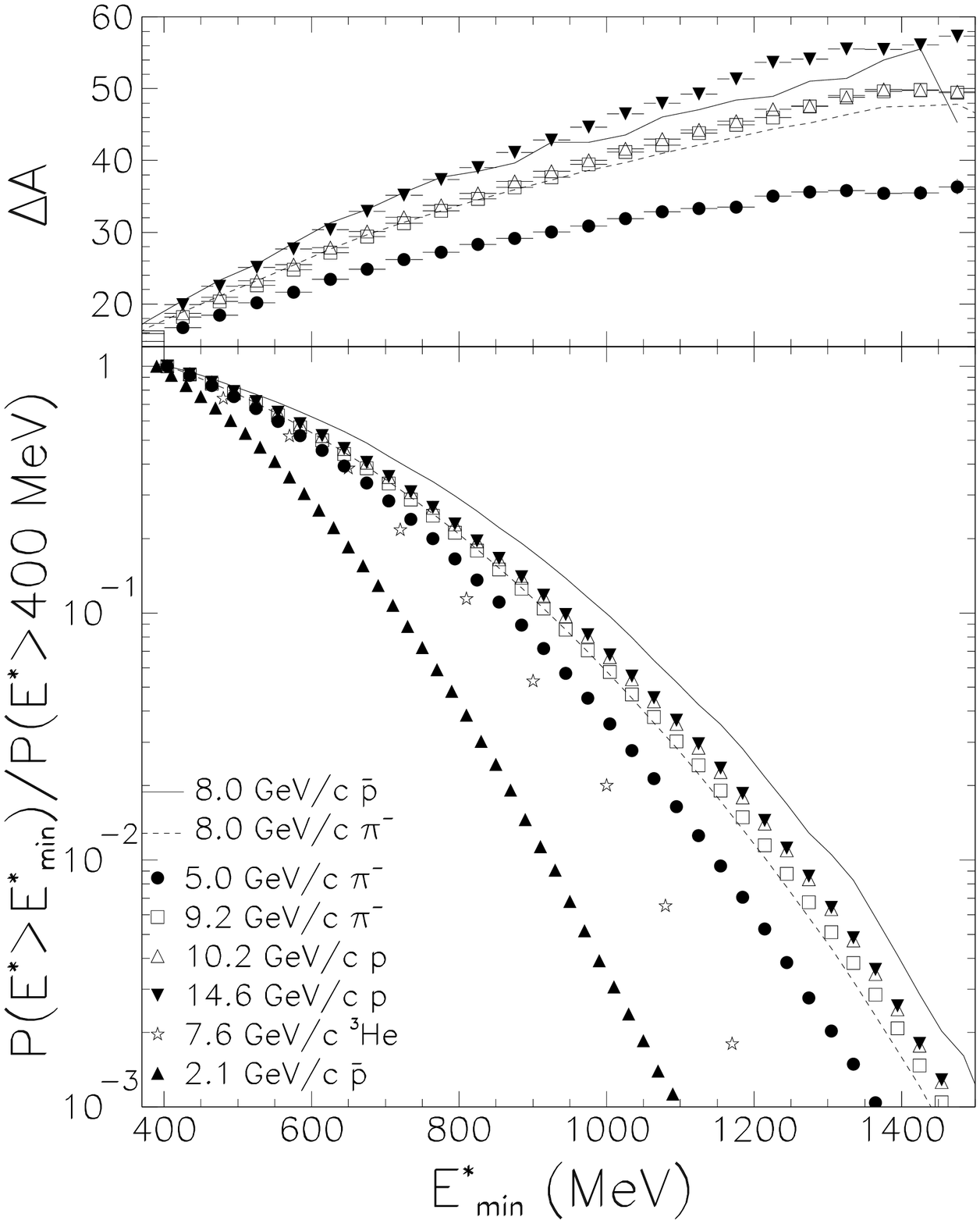,height=8.0in,width=6.5in}}
\vspace{0.25in}
\center{Fig. 3: L. Beaulieu {\em et al.}}
\end{figure}

\newpage

\begin{figure}
\centerline{\psfig{file=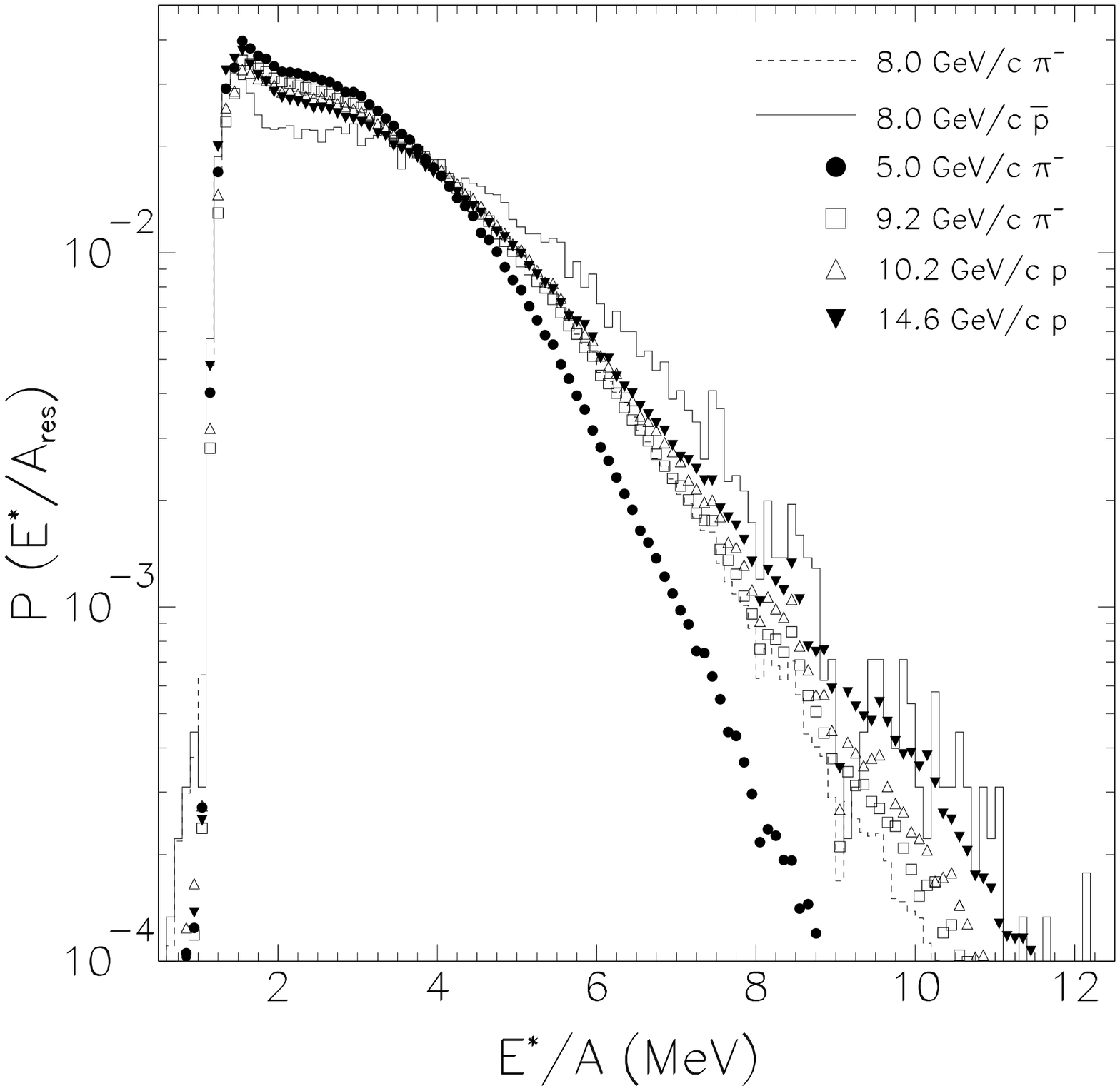,width=7.5in}}
\vspace{1in}
\center{Fig. 4.: L. Beaulieu {\em et al.}}
\end{figure}

\newpage

\begin{figure}
\centerline{\psfig{file=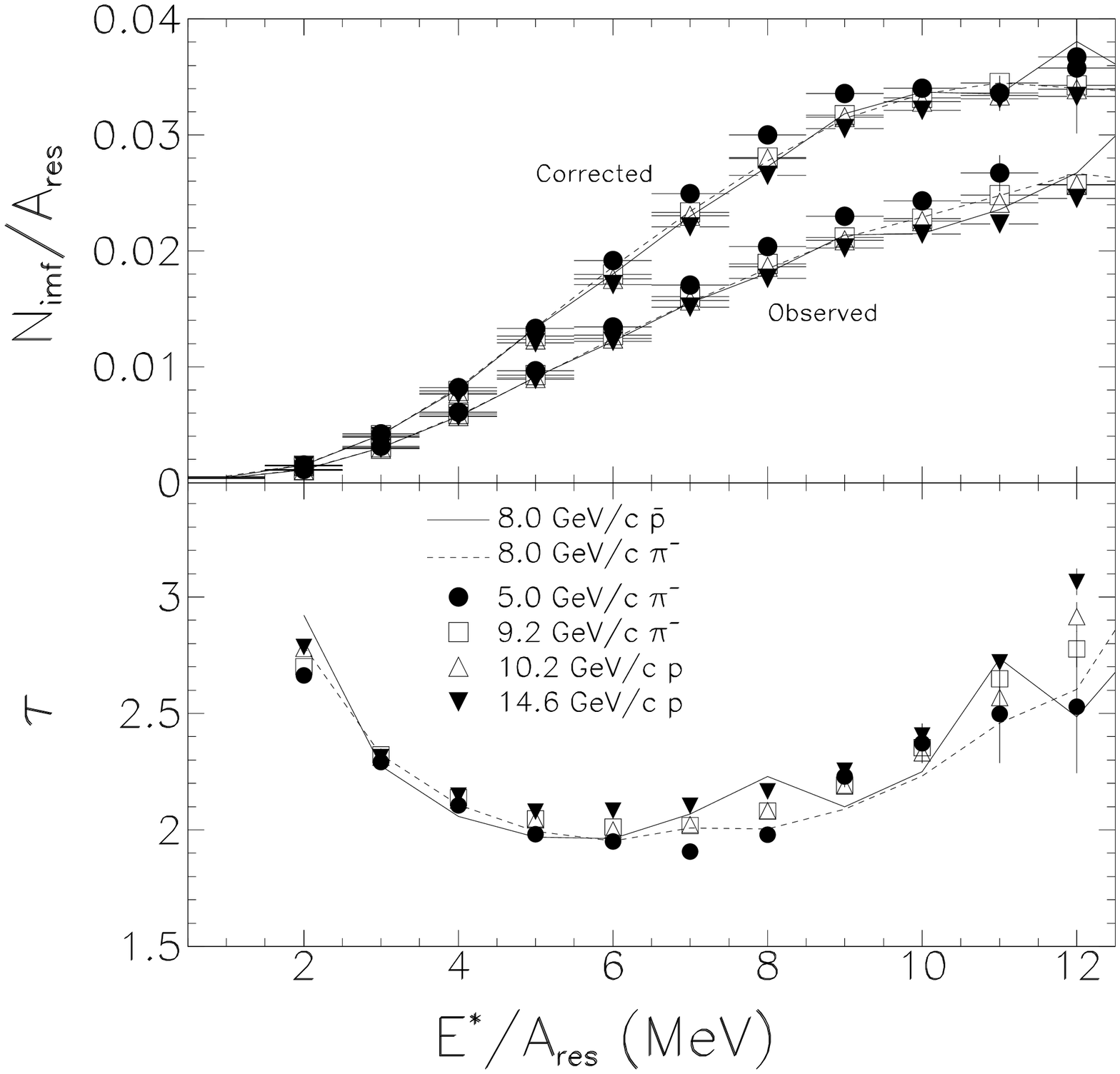,width=7.5in}}
\vspace{1in}
\center{Fig. 5.: L. Beaulieu {\em et al.}}
\end{figure}

\end{document}